\documentclass[superscriptaddress,amssymb,amsmath,twocolumn]{revtex4-1}

\usepackage{color}
\usepackage{amsmath}
\usepackage{amssymb}
\usepackage{mathrsfs}

\usepackage{graphicx}
\usepackage{bm}

\usepackage{multirow}
\usepackage{array}

\renewcommand{\vec}[1]{\bm{#1}}
\newcommand{\derd}[2]{\frac{\mathrm{d} #1}{\mathrm{d} #2}}

\newcommand{\moy}[2]{\langle #1 \rangle_{{#2}}}
\newcommand{\fluct}[1]{#1_\ss'}
\newcommand{\solfrac}{\varphi}
\newcommand{\brangle}{\psi}

\newcommand{\vol}{\Omega}
\newcommand{\vs}{\Omega_\mathrm{s}}
\newcommand{\ex}{\vec{\mathrm{e}}_\mathrm{x}}

\newcommand{\ez}{\vec{\mathrm{e}}_\mathrm{z}}

\renewcommand{\ss}{s}
\newcommand{\car}{\mathcal{C}}
\newcommand{\dint}{\mathrm{d}}
\newcommand{\fem}{FEM}
\newcommand{\crossdim}{D_\ss}
\newcommand{\fields}{\{\vec{t}_\ss,\solfrac,\crossdim\}}
\newcommand{\param}{(\lambda,\beta,\brangle)}
\newcommand{\cd}{C_D}

\begin{document}

\title{A space-averaged model of branched structures}

\author{Diego~Lopez}
\altaffiliation{Corresponding author at: Aix Marseille Universit\'e, CNRS, IUSTI UMR 7343, 13453 Marseille, France}
\email{diego.lopez@univ-amu.fr}
\affiliation{Department of Mechanics, LadHyX  Ecole Polytechnique-CNRS, 91128 Palaiseau, France}
\author{Emmanuel~de~Langre}
\email{delangre@ladhyx.polytechnique.fr}
\affiliation{Department of Mechanics, LadHyX  Ecole Polytechnique-CNRS, 91128 Palaiseau, France}
\author{S\'ebastien~Michelin}
\email{sebastien.michelin@ladhyx.polytechnique.fr}
\affiliation{Department of Mechanics, LadHyX  Ecole Polytechnique-CNRS, 91128 Palaiseau, France}

\begin{abstract}

Many biological systems and artificial structures are ramified, and present a high geometric complexity. In this work, we propose a space-averaged model of branched systems for conservation laws. From a one-dimensional description of the system, we show that the space-averaged problem is also one-dimensional, represented by characteristic curves, defined as streamlines of the space-averaged branch directions. The geometric complexity is then captured firstly by the characteristic curves, and secondly by an additional forcing term in the equations. This model is then applied to mass balance in a pipe network and momentum balance in a tree under wind loading.

\end{abstract}


\maketitle

%

\section{Introduction}
\label{sec:Introduction}

Branched systems are ubiquitous in nature and man-made structures. 
In biological systems, ramification is a mean for increasing exchange surfaces at a given mass; this is commonly observed in blood circulation, pulmonary system \citep{grotberg_1994}, and plants like trees and bushes \citep{niklas_1994}, to list a few. In these various systems, an accurate modeling of fundamental conservation laws is of crucial importance, be it for medical purposes, ecological applications or predictions of mechanical failure.

Over the past decades, numerous studies helped uncover the flow kinematics in the blood system, for instance for targeted drug delivery \citep{tan_2013}, or in lungs for finding geometries {that maximize ventilation in a limited time} \citep{florens_2011a,florens_2011b} and {to} study the behavior of liquid plugs \citep{song_2011,vertti_2012}. In plants, various studies {have been designed to understand} the static and dynamic response to external flows \citep{rodriguez_2008,leung_2011,lopez_2011}.
The modeling complexity of such systems comes from the multiple ramifications and branching points. In these branched systems, robust models exist for individual segments, but branched systems are not easily modeled and often require heavy computations. A key issue is to find a continuous way for modeling these geometries.

A typical example is that of trees submitted to external flows. A  continuous representation of a tree as a tapered beam was proposed by McMahon for analyzing the mechanical stability of a tree under its own weight \cite{macmahon_1975}. This model captures efficiently some key geometric features of tree-like structures and allow for computing accurately the wind-induced loads on an isolated plant \citep{lopez_2011,eloy_2011}. However, this continuous approach does not account for the changes in branch orientation, and the tree effect on the flow cannot be modeled inside the tree crown.
To overcome this issue, many models are based on fractal models for trees \citep{gardiner_1998, endalew_2009, bai_2012}. Such models rely on costly computations and a large number of parameters. Moreover, despite the variety of existing models, there is a lack of a general formulation of conservation laws in branched systems.

\begin{figure*}[tb]
	\centering
		\includegraphics[width = 1\textwidth]{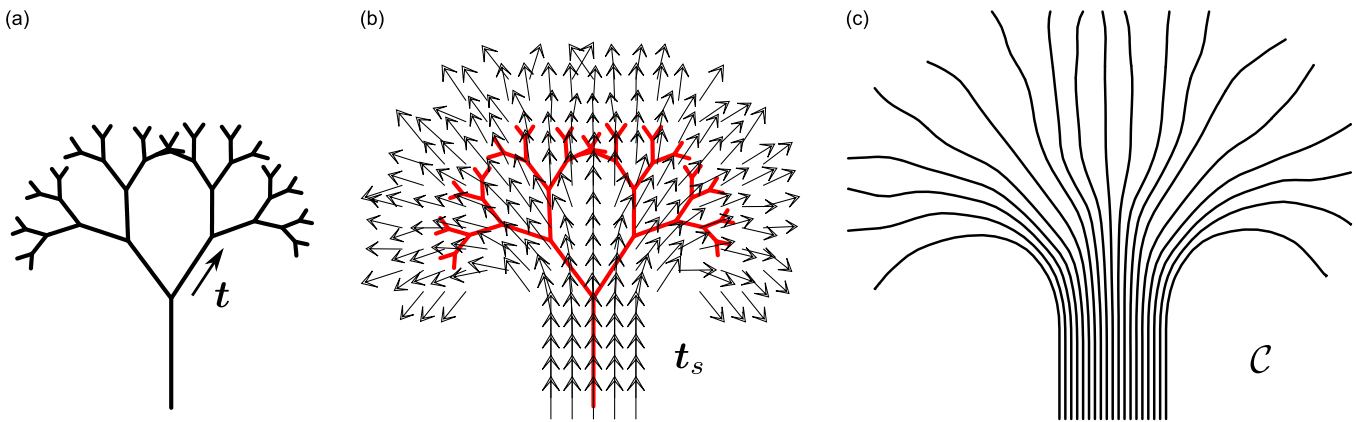}
	\caption{Space-averaged branching model: (a) ramified system with oriented branches ($\vec{t}$), (b) Volume averaged branch direction $\vec{t}_\ss$, (c) characteristic curves $\car$ equivalent to the branched system in SAB model.}
	\label{fig:1}
\end{figure*}

In this paper, we present a new model for space-averaged branching (SAB) in conservation laws. The purpose of this work is to provide a continuous formulation of conservation laws in branched systems, represented by a small number of parameters and applicable to a large variety of problems{, in particular for solving full fluid-structure computations on branched systems through a porous medium approach. More specifically, we expect that the proposed approach will help in modeling complex structures involving large number of branching, avoiding the fine description of each and every segment}. 
The present SAB model is inspired from homogenization techniques and porous media approach. 
We obtain an equivalent problem where a branched system is represented by independent characteristic curves, on which specific conservation equations are solved. These characteristic curves correspond to streamlines of the average branch direction, as sketched in Fig.~\ref{fig:1}. 
The SAB model is derived in Sec.~\ref{sec:TreeLikeStructureVolumeDescription}. We present then two applications of the model, first on a case study of flow rate computation in a simple pipe network in Sec.~\ref{sec:ModelValidation}, and then on the problem of trees submitted to an external flow in Sec.~\ref{sec:SABForWindInducedLoadsOnATree}. Finally, a general discussion and conclusion is given in Sec.~\ref{sec:ConclusionAndDiscussion}.

\section{Space-averaged branching model}
\label{sec:TreeLikeStructureVolumeDescription}

\subsection{Definitions and problem equations}
\label{sec:FundamentalEquationsAndDefinitions}

We consider a branched system where the segments between two branching points are oriented and described by the corresponding tangent vector $\vec{t}$ (Fig.~\ref{fig:1}a). The segments are slender, so that a segment length is much larger than its transverse dimension. The resulting description is thus one-dimensional along the segments.
Under these assumptions, a general formulation of the conservation of a vectorial quantity $\vec{Q}$ along the system is
\begin{align}
	\derd{\vec{Q}}{x} + \vec{G}(x) = 0, \quad \sum{\vec{Q}^-} = \sum{\vec{Q}^+},
	\label{eq:general}
\end{align}
where $\vec{G}$ is a forcing term, $x$ the curvilinear coordinate, and we use superscript $-$ (resp.~$+$) to characterize a segment oriented towards (resp.~away from) the branching point, according to the system orientation given by the tangent vector $\vec{t}$ (Fig.~\ref{fig:2}). 
The first relation is the conservation of $\vec{Q}$ along a segment, and the second one gives a relation at the nodes of the structure.
Such conservation equations are ubiquitous in branched systems, and their complexity {arises from} the discontinuities introduced by branching nodes. 
In the general case, Eq.~\eqref{eq:general} is a vectorial equation, but can be decomposed into a set of scalar equations by projection on a fixed frame. In the following, we derive the SAB model considering a scalar problem, $\mathrm{d}Q/\mathrm{d}x + G(x) = 0$.

Whereas the initial branched system has no volume (1D description), it is necessary to introduce its finite volume for averaging purposes (see Fig.~\ref{fig:2}a). 
In order to obtain space-averaged quantities, we introduce a representative volume $\vol$ and denote $\vs$ the volume occupied by the branched system included in  $\vol$, as sketched in Fig.~\ref{fig:2}a. We define the volume fraction $\solfrac = \vs/\vol$. The representative volume $\vol$ must be large compared to the typical diameter of the branched system's segments \citep{whitaker_1999}.
We use a standard space average operator over the branched system, noted $\moy{.}{\ss}$, 
\begin{equation}
	\moy{.}{\ss} =  \frac{1}{\vs} \int_{\vs}{. \dint \vol},
	\label{eq:moy}
\end{equation}
for a quantity $Q$ defined in the system. This formalism is typically used in porous media analysis, where $\vs$ stands for the volume occupied by a solid and $\vol - \vs$ is occupied by a fluid \citep{pedras_2001, hoffman_2004}.

\subsection{Volume equation derivation}
\label{sec:MechanicalModelAndAveragingMethod}
For any quantity $Q(x)$, {where $x$ is the curvilinear coordinate along the segment}, we introduce in the volume $\vs$ a continuously differentiable function $q$ corresponding to $Q$ per unit section, so that in a cross section normal to $\vec{t}$,
\begin{equation}
	q = q(x) =\frac{Q(x)}{A(x)},
	\label{eq:qdef}
\end{equation}
{where $A(x)$ is the local cross-section.}
This definition yields some singularities at the branching points and at the borders of the averaging volume. Due to the high slenderness of the segments, these singularities are easily overcome without loss of generality; these technical points are discussed in \ref{sec:app_hg}.

\begin{figure}[tb]
	\centering
		\includegraphics{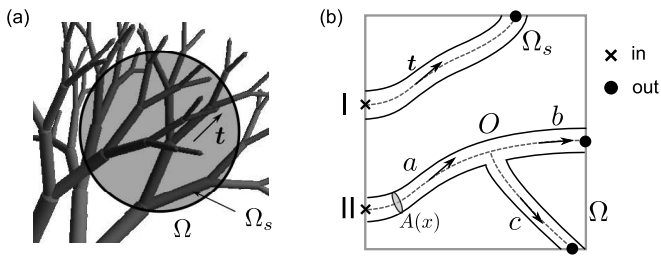}
	\caption{(a) Space averaging domain; (b) Example for the averaging method and corresponding notations.}
	\label{fig:2}
\end{figure}

We consider the sketch of Fig.~\ref{fig:2}b for obtaining volume equations. We denote $Q^\text{in}$ (resp.~ $Q^\text{out}$) the sum of $Q$ where the segments go into $\vol$ (resp.~out of $\vol$) with respect to $\vec{t}$. According to the previous notations and slenderness hypothesis, we can write for segment I in Fig.~\ref{fig:2}b
\begin{equation}
	Q_{I}^\text{out} - Q_I^\text{in} = \oint_{\partial {\vs}_I}{\frac{Q}{A}\vec{t}.\vec{n}_{\vs} \dint \mathcal{S}} = \oint_{\partial {\vs}_I}{q\vec{t}.\vec{n}_{\vs} \dint \mathcal{S}},
	\label{eq:slendernesshyp}
\end{equation}
where $\partial{\vs}_I$ is the border of segment I.
Since $q$ is continuously differentiable in $\vs$, we can apply the divergence theorem. We can then introduce the space-average operator as defined in Eq.~\eqref{eq:moy}, {and we use a special property for the volume average of the spatial divergence{, noted $\nabla.$,}
\begin{equation}
	\solfrac\moy{\nabla. q\vec{t}}{\ss} = \nabla. \left(\solfrac\moy{q\vec{t}}{\ss}\right) + \frac{\solfrac}{\vs} \int_{\partial {\vs}}{q\vec{t}.\vec{n} \dint \mathcal{S}},
	\label{eq:nabla}
\end{equation}
where $\partial {\vs}$ the interface between $\vol$ and $\vs$, and $\vec{n}$ the normal to the interface oriented towards $\vs$ \citep{whitaker_1999}.}
As a result, the sum over each independent segment (here noted I and II) gives
\begin{equation}
	Q^\text{out} - Q^\text{in} = \frac{\vs}{\solfrac} \nabla.\left(\solfrac\moy{q\vec{t}}{\ss}\right).
	\label{eq:qinout1}
\end{equation}

We consider now the conservation equations given in Eq. \eqref{eq:general}, which give, for segment I (or II) of Fig.~\ref{fig:2}b,
\begin{equation}
	Q_I^\text{out} - Q_I^\text{in} = -\int_{\text{in}_I}^{\text{out}_I}{G \dint x} = -\int_{{\vs}_I} {g \dint \vol}.
\end{equation}
The same analysis can be done on segment II using the conservation of $Q$ at a branching point, leading in the general case to
\begin{equation}
	Q^\text{out} - Q^\text{in} = -\int_\text{in}^\text{out}{G \dint x},
\end{equation}
where the integration from ``in'' to ``out'' represents the summation along every oriented segment in $\vol$. By neglecting the effect of the branching region (see \ref{sec:app_hg}), we can write in the general case
\begin{equation}
	Q^\text{out} - Q^\text{in} = - \vs \moy{g}{\ss}.
	\label{eq:qinout2}
\end{equation}

Combining Eqs.~\eqref{eq:qinout1} and \eqref{eq:qinout2}, we get the volume equation on $q$, resulting from Eq.~\eqref{eq:general} for any scalar quantity $Q$,
\begin{equation}
	\nabla.\left(\solfrac\moy{q\vec{t}}{\ss}\right) + \solfrac \moy{g}{\ss} = 0.
	\label{eq:generalvolume}
\end{equation}
This equation was obtained using only the high slenderness of the segments constituting the branched system (see details in \ref{sec:app_hg}). At this point, Eq.~\eqref{eq:generalvolume} is a general volume formulation resulting from a one-dimensional description {similar to that} of Eq.~\eqref{eq:general}, but its solution is no straightforward. In the next section, we present a first order approximation of the model equation.

\subsection{First order approximation}
\label{sec:ModelAssumption}

Eq.~\eqref{eq:generalvolume} can be decomposed using spatial fluctuations $\fluct{Q} = Q - Q_\ss$, with a simplified notation $Q_\ss=\moy{Q}{\ss}$. The resulting equation reads then
\begin{align}
	\nabla.\left(\solfrac q_\ss \vec{t}_\ss\right) + \solfrac g_\ss + \nabla.\left(\solfrac\moy{\fluct{q}\fluct{\vec{t}}}{\ss}\right) = 0
	\label{eq:volfluct}
\end{align}

\begin{figure}[tb]
	\centering
		\includegraphics{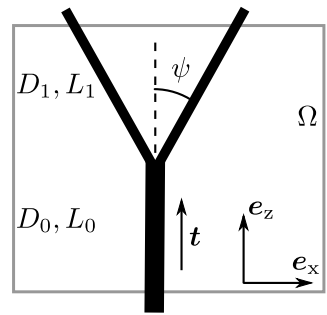}
	\caption{Example of averaging over a Y-shaped geometry: effect of first order approximation on tangent vector $\vec{t}$.}
	\label{fig:3}
\end{figure}

We consider here as a first order approximation that the fluctuations, and their derivatives, are negligible compared to the corresponding mean values,
\begin{equation}
	|\fluct{Q}| \ll |Q_\ss|.
	\label{eq:hypo}
\end{equation}
Equivalently, this consists in modeling any quantity by its space-averaged value in the branched system volume, $Q \equiv Q_\ss$.
Such assumption implies that the variations of any quantity defined in the real structure have to remain relatively small. In particular, the changes in orientation should not be too important. For instance, if we consider a Y-shaped branch with branching angle $\brangle$ (see Fig.~\ref{fig:3}), the solid volume average of $\vec{t}$ reads
\begin{equation}
	\vec{t}_\ss = \frac{D_0^2L_0 + 2D_1^2 L_1 \cos \brangle}{D_0^2L_0 + 2D_1^2 L_1} \ez,
	\label{eq:expleT}
\end{equation}
whereas $\vec{t}$ is equal to $\ez$ in the first level, and then $\cos\brangle \ez \pm \sin \brangle \ex${, the $\ex$ and $\ez$ vectors being defined in Fig.~\ref{fig:3}}. When the angle $\brangle$ is small, this hypothesis is justified, and it is exact when $\brangle =0$. This provides a range of validity of the first order approximation of Eq.~\eqref{eq:hypo}.

Under the assumption of Eq.~\eqref{eq:hypo}, a new form of the volume equation is found, neglecting the last term of Eq.~\eqref{eq:volfluct},
\begin{align}
	\nabla.\left(\solfrac q_\ss \vec{t}_\ss\right) + \solfrac g_\ss = 0.
	\label{eq:modQ}
\end{align} 
Here, one only needs the average branch direction field $\vec{t}_\ss$, the volume fraction $\solfrac$ and the average forcing field $g_\ss$, where
\begin{equation}
	\vec{t}_\ss =  \frac{1}{\vs} \int_{\vs}{\vec{t} \dint \vol}, \quad \solfrac=\frac \vs\vol,\quad g_\ss =  \frac{1}{\vs} \int_{\vs}{g \dint \vol}.
	\label{eq:SABfields}
\end{equation}
The volume equation obtained above can be solved directly, however a particular solution technique is presented in the next section.

\subsection{Model equations: solution on characteristic curves}
\label{sec:HomogenizedModelEquations}

\begin{figure}[tb]
	\centering
		\includegraphics{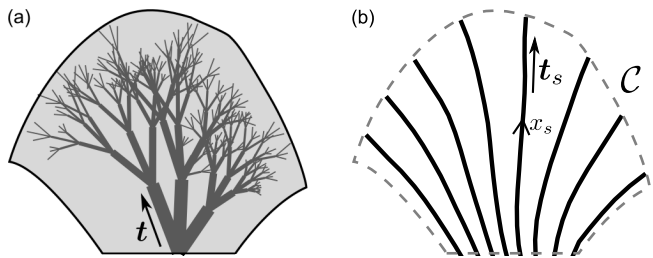}
	\caption{Space-averaged branching model: (a) branched system and domain for space averaging; (b) SAB equivalent medium, described by characteristic curves $\car$.}
	\label{fig:4}
\end{figure}

By expending the first term of Eq.~\eqref{eq:modQ}, we get
\begin{equation}
	\nabla.\left(\solfrac\moy{q\vec{t}}{\ss}\right) = \solfrac\vec{t}_\ss.\nabla q_\ss + q_\ss\nabla.(\solfrac\vec{t}_\ss).
\end{equation}
The term $(\vec{t}_\ss.\nabla)$ corresponds to a derivative along a curve tangent to $\vec{t}_\ss$. For a given streamline $\car$ of the space-averaged branch direction field $\vec{t}_s$, we define $x_s$ the space-averaged curvilinear coordinate along $\car$. We can write 
\begin{align}
	\vec{t}_\ss.\nabla = \derd{}{x_s} \quad \text{on } \car.
\end{align}
The volume equation \eqref{eq:modQ} can therefore be written as a set of one-dimensional equations along characteristic curves $\car$ 
\begin{align}
	\solfrac|\vec{t}_\ss|\derd{q_\ss}{x_\ss} + q_\ss\nabla.(\solfrac\vec{t}_\ss) + \solfrac g_\ss = 0,
	\label{eq:mod1dq}
\end{align}
where $x_s$ is the space-averaged curvilinear coordinate along each curve $\car$.
The characteristic curves correspond to streamlines of the average branch direction field $\vec{t}_\ss$, as sketched in Fig.~\ref{fig:4}.
It is interesting to note that the resulting equations of the SAB model are one-dimensional, as {for} the initial equation Eq.~\eqref{eq:general}, but with the advantage of being now defined everywhere continuously.
 
The SAB equation \eqref{eq:mod1dq} contains an additional term, namely $\left(\nabla.\solfrac\vec{t}_\ss\right)$. This term is crucial in the SAB formulation, as it contains branching effects and changes in diameter or orientation. This branching term vanishes for cylindrical sections, but is non-zero when the diameter changes or branching occurs. 
This term acts as an additional forcing, and will be referred to in the following as geometrical forcing.
Indeed, even in the absence of external forcing,  segment size variations and branching can cause variations of $Q$. This will be illustrated in the next section. The geometrical forcing $\left(\nabla.\solfrac\vec{t}_\ss\right)$ therefore holds the geometric complexity characterizing the real system.

The SAB continuous medium equivalent to a branched system is thus modeled by a set of characteristic curves as shown in Fig.~\ref{fig:4}b, on which particular one-dimensional equations are solved. The geometry is thus characterized only by the vector field $\vec{t}_\ss$ and the volume fraction $\solfrac$ of the branched system with respect to a reference volume. In the next two sections, we show applications of the SAB model, first to mass balance, then to momentum balance.

\section{Application to mass balance}
\label{sec:ModelValidation}

\begin{figure}[tb]
	\centering
		\includegraphics{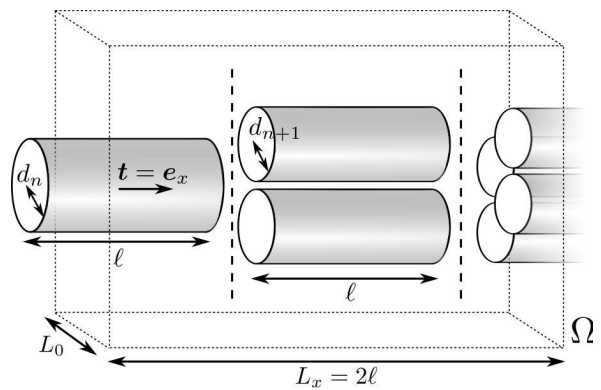}
	\caption{Example of a 1D pipe network, where each pipe divides into two pipes after a length equal to $\ell$. The averaging volume $\vol$ is shown in dotted lines.}
	\label{fig:5}
\end{figure}

We consider mass conservation in a one-dimensional pipe network, as sketched in Fig.~\ref{fig:5}.
This system is made of cylindrical pipes with fixed length $\ell$ and possibly varying diameter $d_n$ at level $n$. At a branching point, each pipe divides into two pipes, and we neglect the branching regions.
We consider an incompressible fluid flowing at a constant flow rate in the system along direction $\ex$, without any flow source or sink. 
The averaging volume is such that its length in the pipe direction is $L_x=2\ell$; the perpendicular dimension $L_0$ (in $y$ and $z$ directions) is taken to be arbitrarily large, so that the SAB problem is one-dimensional. 
Since the network is oriented along $\ex$, the space-averaged segment direction is  $\vec{t}_s=\ex$.
The problem can therefore be modeled by a single characteristic curve $\car$, along the $x$-axis.

We focus here on the flow rate $Q$. As a result, the quantity $q$ defined as $Q$ per unit section is in fact the fluid velocity $U$ averaged over a pipe section. In the absence of external source or sink, the SAB equation for the flow velocity reads
\begin{align}
	\solfrac\derd{U_\ss}{x_\ss} + U_\ss\derd{\solfrac}{x_\ss} = 0,
\end{align}
where $U_\ss$ is the space-averaged velocity.  
We first consider the case where the total section is conserved at a branching point, $2d^2_{n+1}=d_n^2$; this corresponds to the constant velocity case. The volume fraction $\solfrac$ occupied by pipes is constant, and the SAB model equation reads
\begin{align}
	\solfrac\derd{U_\ss}{x_\ss} = 0,
\end{align}
which naturally results in a constant velocity.

\begin{figure}[tb]
	\centering
		\includegraphics{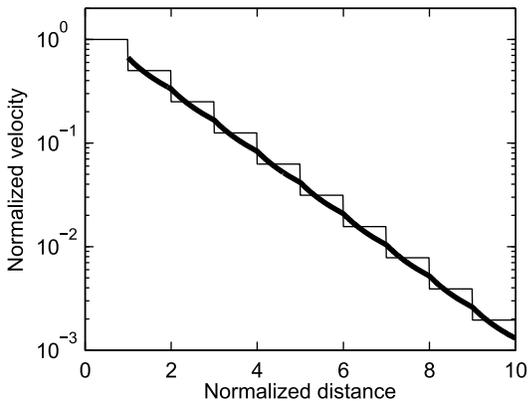}
	\caption{Normalized flow velocity along a 1D pipe network of identical pipes, using direct analysis (thin line, $U/U_0 = f(x/\ell)$) and SAB model (thick line, $U_s/U_0 = f(x_s/\ell)$).}
	\label{fig:6}
\end{figure}

We consider now that all segments are identical, $d_{n+1}=d_n$. The velocity is reduced at each branching point by a factor 2, and reads $U(x) = U_0 2^{-\lfloor x/\ell\rfloor}$, where $\lfloor.\rfloor$ is the floor function. 
{Noting $x_\ss$ the position of the center of  the averaging volume, the number of incoming pipes in $\vol$ is equal to $2^{\lfloor {x_\ss}/\ell\rfloor-1}$. It is then possible to compute directly the volume fraction $\solfrac$ which is not constant in this case}
\begin{align}
	\solfrac(x_\ss) = \frac{3\pi}{8}\frac{d_0^2}{L_0^2}\left(1+\frac {x_\ss}\ell - \left\lfloor\frac {x_\ss}\ell\right\rfloor\right)2^{\lfloor {x_\ss}/\ell\rfloor-1}.
\end{align}
The resulting equation for the velocity in the SAB model reads therefore
\begin{align}
	\ell\left(1+\frac {x_\ss}\ell - \left\lfloor\frac {x_\ss}\ell\right\rfloor\right)\derd{U_\ss}{x_\ss} + U_\ss= 0, \quad U_\ss(x_\ss = \ell) = 2 U_0/3,
\end{align}
which can be solved numerically.
The evolution of the velocity is plotted in Fig.~\ref{fig:6}, showing  a perfect agreement between SAB and direct computation. 
This example shows how crucial the geometrical forcing $\left(\nabla.\solfrac\vec{t}_\ss\right)$ is in the SAB model. In this case, it is in fact the only forcing term, accounting for the increasing total cross-section.

These simple case was derived as a first validation of the SAB model in a problem with an analytical solution. In particular, the number of characteristic curves has no influence here, since this problem is modeled by a single characteristic curve. In the next section we turn to a more elaborate problem for a complete validation of the model.

\section{Application to momentum balance}
\label{sec:SABForWindInducedLoadsOnATree}

\subsection{SAB equations of wind-induced loads on a tree}
\label{sec:MechanicalAnalysis}

The space-averaged branching model is used for computing the loads induced by a steady wind on a standing tree-like structure. The model can be developed in the general three-dimensional case, however we present hereafter only a two-dimensional example, for the sake of clarity.
This case is inspired from \cite{lopez_2011}.
We consider a two-dimensional sympodial tree made of cylindrical branches, described by three parameters: (i) the branching ratio $\lambda$, giving the reduction of section through branching ($\lambda = (D^+/D^-)^2$, $\lambda <1$), (ii) the slenderness exponent $\beta$, giving the relationship for length and diameter evolution in branch segments of the tree ($D^+/D^-=(L^+/L^-)^\beta$, $1<\beta<2$), and (iii) the branching angle $\brangle$ as defined in Fig.~\ref{fig:3}. The number of branches emerging from one branch at a branching point is typically equal to $1/\lambda$ \citep{rodriguez_2008, lopez_2011}.

\begin{figure}[tb]
	\centering
		\includegraphics{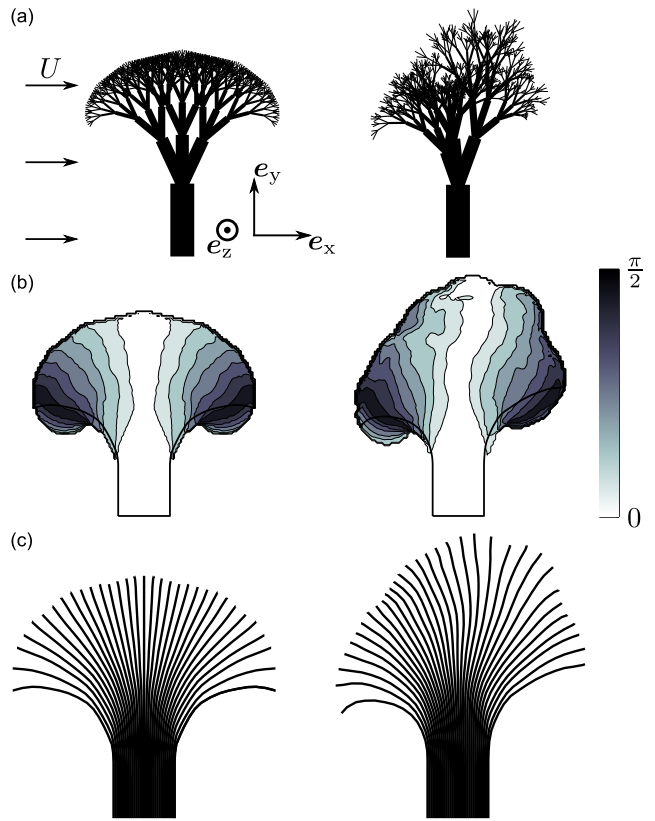}
	\caption{Symmetric and random {tree-like structures} used for the validation of the homogenized model: (a) Real structure under cross flow, (b) average branch angle with the vertical axis (absolute values), (c) characteristic curves used in the SAB model.}
	\label{fig:7}
\end{figure}

We consider here two geometries: (i) a symmetric tree with $\lambda_0 = 0.3$, $\beta_0=1.5$ and $\brangle_0 = 25^\circ$, and (ii) a tree with random variations of its parameters $\param$ of 30\% around the default values $\lambda_0$, $\beta_0$ and $\brangle_0$. These two geometries are shown in Fig.~\ref{fig:7}, along with contour values of the average branch angle with respect to the vertical axis, and the corresponding characteristic curves used in the SAB model.

The structure is submitted to an external flow, and is held by a perfect clamping at the base, the top being free of loads (Fig.~\ref{fig:7}a). 
We assume hereafter that the deformations induced by the flow are negligible. The equations that govern the mechanical response of this structure under an external flow read
\begin{align}
	\label{eq:eulerb}
	&\derd{\vec{V}}{x} + \vec{F} = \vec{0}, \quad \derd{\vec{M}}{x} + \vec{t}\wedge \vec{V} = \vec{0}, \\
	&\vec{F} = \frac{1}{2} \rho \cd D |\vec{U}.\vec{n}|(\vec{U}.\vec{n})\vec{n}, \nonumber
\end{align}
where $\vec{V}$ is the sum of the normal and shear forces, and $\vec{M}$ is the bending moment in 2D \citep{salencon_2001}.  $\vec{F}$ the fluid force, resulting from a normal pressure drag oriented along the normal to the branch axis $\vec{n}$; $\rho$ is the fluid density, $\vec{U}$ the fluid velocity and $\cd$ the drag coefficient, taken to be 1 \citep{lopez_2011}. In this problem, there is no external {torque} applied on the structure. 
As the problem is isostatic and the configuration known in the absence of deformations, Eq.~\eqref{eq:eulerb} describes completely the problem. 

These equations correspond to the general form of conservation laws given in Eq.~\eqref{eq:general}.
Applying the procedure described in Sec.~\ref{sec:TreeLikeStructureVolumeDescription} on each component of Eq.~\eqref{eq:eulerb}, we derive the equations of the SAB model on characteristic curves, 
\begin{align}
	\label{eq:mod1dv}
	& \solfrac|\vec{t}_\ss|\derd{\vec{v}_\ss}{x_\ss} + \vec{v}_\ss\nabla.\solfrac\vec{t}_\ss + \solfrac \vec{f}_\ss = \vec{0}, \\
	\label{eq:mod1dm}
	& \solfrac|\vec{t}_\ss|\derd{\vec{m}_\ss}{x_\ss} + \vec{m}_\ss\nabla.\solfrac\vec{t}_\ss + \solfrac\vec{t}_\ss\wedge\vec{v}_\ss = \vec{0},
\end{align}
using the same notations for lower-case letters.
The external flow is taken to be uniform, $\vec{U} = U\ex$ (Fig.~\ref{fig:7}a). Using the assumption of Eq.~\eqref{eq:hypo}, the space-averaged forcing reads
\begin{equation}
	\vec{f}_\ss \equiv \frac{1}{2}\rho \cd {U}^2 |\ex.\vec{n}_\ss|(\ex.\vec{n}_\ss)\vec{n}_\ss,
	\label{eq:forcemodel}
\end{equation}
where $\vec{n}_\ss = \ez \wedge \vec{t}_\ss$ in 2D. 
Note that  this analysis is carried out in the two-dimensional case, therefore the cross section $A$ is equal to the branch diameter $D$.
Finally, we introduce the maximum bending stress in a branch section, defined as $\Sigma = 32M/\pi D^3$ \citep{niklas_1992, lopez_2011}. {Noting that $m_\ss=\moy{M/D}{\ss}$,} the bending stress is modeled here by
\begin{equation}
	\Sigma \equiv \Sigma_s = \frac{32}{\pi}\frac{m_\ss}{\crossdim^2}.
	\label{eq:sigma}
\end{equation}

The present SAB model will be compared to finite element computations on exact geometries, noted \fem, using a standard finite element software (CASTEM v.3 M, \cite{verpeaux_1988}). 
{The finite-element model consists of Euler-Bernoulli beam elements, each branch being described by ten mesh-elements. This refinement was found sufficient to compute the static loads according to a convergence test. The bottom boundary condition is enforced by preventing all displacement and rotation of the anchoring node, and the computations are carried out in three dimensions for this 2D geometry. This software was previously used for computing loads on tree-like structures \cite{rodriguez_2008,theckes_2011}.}
The FEM method solves directly Eq.~\eqref{eq:eulerb} and the bending stress is then computed using $\Sigma = 32M/\pi D^3$, whereas the SAB computation solves Eqs.~\eqref{eq:mod1dv}-\eqref{eq:sigma}.
For the SAB model, the fields $\fields$ are obtained numerically, with a representative averaging surface (in two dimensions) whose typical dimension is of the order of the length of the first branches after the trunk. Applying the averaging method along the trunk can seem dubious, but we want here to keep the model free of a matching condition between the trunk and the tree crown; the same model is therefore used throughout the entire structure.
The model equations, Eqs.~\eqref{eq:mod1dv} and \eqref{eq:mod1dm}, are then solved using an implicit Euler method on each characteristic curve $\car$.

\subsection{Flow-induced loads}
\label{sec:FlowInducedLoadsOnSympodialTrees}

We compute the loads induced by a uniform cross flow on {an idealized tree}, using the present space-averaged branching model and finite elements computations. Note that the loads are proportional to the fluid dynamic pressure $\rho U^2$, hence one only needs to compute the loads for a {single fluid velocity $U_0$. 
The forces are then normalized by $V_0 = \frac 12 \rho \cd U_0^2 D_0 L_0$, the moments by $M_0 = V_0 L_0$ and the stresses by $\Sigma_0 = 32 F_0 L_0/\pi D_0^3$, where $D_0$ and $L_0$ are the trunk diameter and length.}
In this problem, the characteristic curves of the SAB representation are independent of each other. As a result, the number of curves only determines the spatial resolution.

\begin{figure*}[tb]
	\centering
		\includegraphics[width = 1\textwidth]{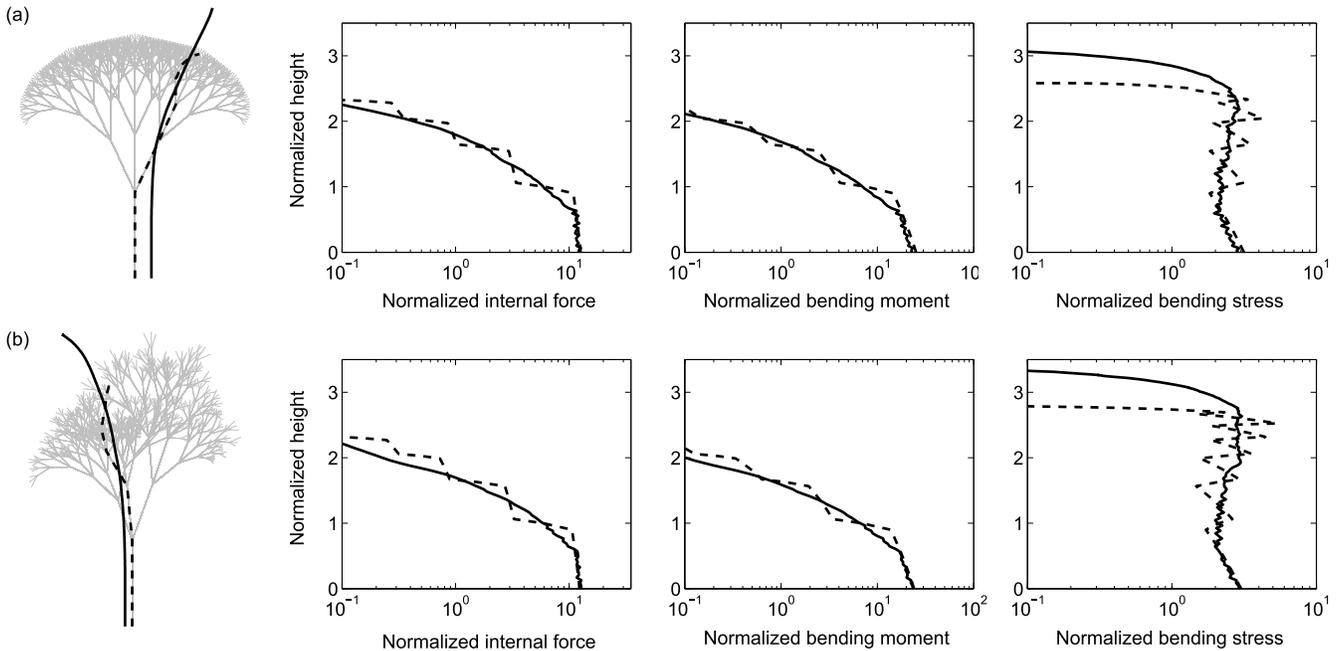}
	\caption{Flow-induced loads on a tree ($\lambda_0 = 0.3, \beta_0 = 1.5, \brangle_0 = 25^\circ$) under uniform cross-flow: (a) symmetric and (b) random tree. The evolution of the normalized internal force $V/V_0$, bending moment $M/M_0$ and bending stress $\Sigma/\Sigma_0$ are computed by finite element method on the actual geometry (dashed lines) and using the SAB model on the equivalent characteristic curve (solid lines). The corresponding branch and characteristic curve are shown using the same line style.}
	\label{fig:8}
\end{figure*}

The results for both symmetric and random trees are shown in Fig.~\ref{fig:8}. 
This figure shows the evolution of the {normalized internal force $V/V_0$, bending moment $M/M_0$ and bending stress $\Sigma/\Sigma_0$} along a typical branch and characteristic curve corresponding to the same physical region. The SAB model shows a very good agreement with the finite element computations, and captures the loads evolution as well as their order of magnitude. Note that there is no fitting parameter in this model.

The discontinuities visible in the \fem~curves are due to the branching points, and are naturally smoothed by the SAB model. These discontinuities are even more significant for the bending stress, as it depends strongly on the branch diameter. For this mechanical quantity, the choice {of} Eq.~\eqref{eq:sigma} has an important influence on the value computed by the SAB model. With the simple choice made here, we observe a good agreement, allowing us to recover the right order of magnitude and stress profile, with a local maximum near the top, i.e.~in the tree crown. 

The space-averaged branching model provides a good estimate of the flow-induced loads, removing the discontinuities introduced by the branching points, without any fitting parameter. Moreover, the computations on a random {tree-like structures} show that there is no particular symmetry required in the initial system.
The present model was derived under the assumption of Eq.~\eqref{eq:hypo}, implying in particular that the branching angle should be small. Whereas this was the case in the example of mass conservation in a 1D pipe system, the branching angle is not negligible in the geometries considered in this section. We see therefore that the assumption of Eq.~\eqref{eq:hypo} is not very restrictive, and the SAB model remains valid for various realistic systems.

\subsection{Application to flow-induced pruning}

We apply now this model to the example of flow-induced pruning in {idealized} trees submitted to wind \citep{lopez_2011}. 
Flow-induced pruning refers to the pruning of branches (or more generally parts of the structure) by the external flow, as the velocity increases, thereby reducing the drag experienced by the tree. In this section, we focus on flow-induced pruning in the absence of deformations.

The pruning procedure is that described in \cite{lopez_2011}: when the velocity increases, breakage occurs where and when the bending stress reaches the yield stress $\Sigma_c$. The broken branch is then removed, and the computation continues on the remaining structure. For the SAB model, the same procedure is applied on each characteristic curve. The fluid velocity is further increased until a new breaking event occurs. 
{The SAB model is not adapted for computing the loads in a single branch, in particular in the tree trunk. In the following,} flow-induced pruning is studied only in the tree crown, and compared with the finite elements computations by looking at the evolution of the total drag at the base of the crown. In the SAB computations, this drag is the average value over the characteristic curves at the base of the tree crown. 
{Finally, in the finite element analysis, the computation stops when the structure is entirely broken. 
This corresponds in fact to the first breaking event located at the base of the structure. We choose similarly to stop the SAB computations when 1\% of the characteristic curves are broken at their base.}

\begin{figure}[tb]
	\centering
		\includegraphics{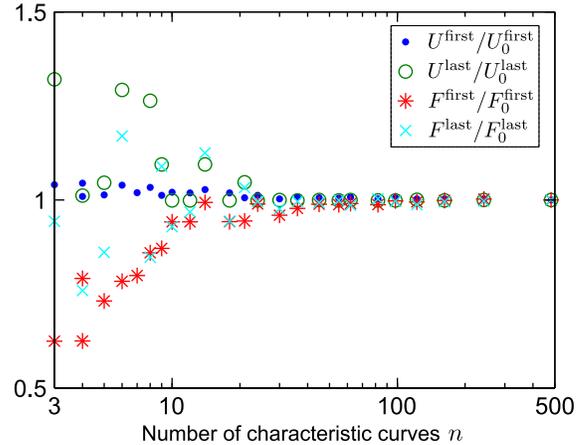}
	\caption{Convergence of the SAB model as the number of characteristic curves $n$ increases. Symbols correspond to velocities at first and last breaking events ($\bullet$, $\circ$) and corresponding drag ($\ast$, $\times$), normalized by their value for $n_0=500$.}
	\label{fig:9}
\end{figure}

This problem is now a global problem involving the  entire domain, and raises the issue of convergence with respect to the number of characteristic curves.
We address this issue by comparing four quantities for different numbers of characteristic curves $n$ to their value for $n_0=500$. Those quantities are the flow velocities at the first and last breaking events ($U^\text{first}$, $U^\text{last}$) and the corresponding drag ($F^\text{first}$, $F^\text{last}$). We see in Fig.~\ref{fig:9} that these quantities vary dramatically at low numbers of characteristic curves, but converge for $n\geq50$. 
It is interesting to note that this convergence occurs at a much smaller number of characteristic curves than the maximum number of parallel branches (here equal to $729$), possibly reducing the computational cost.

\begin{figure}[tb]
	\centering
		\includegraphics{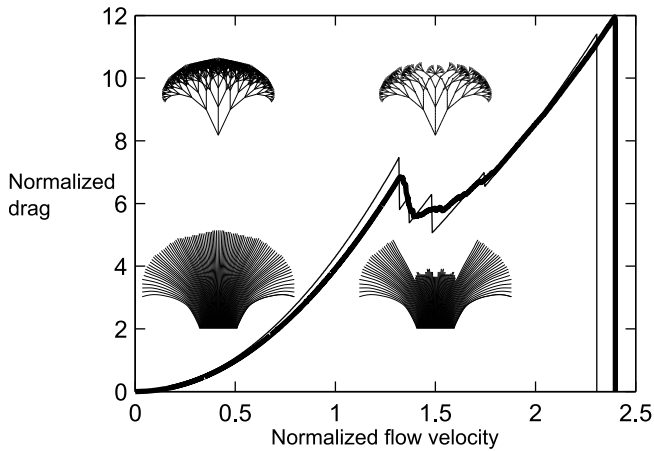}
	\caption{Evolution of the drag under increasing velocity for a symmetric tree undergoing flow-induced pruning, obtained by finite element computations (thin line) and space-averaged branching model (thick line). Two typical geometries are shown, before the first breaking event ($U=0.4$) and after some breaking events ($U =1.5$).}
	\label{fig:10}
\end{figure}

We turn now to the drag evolution as the flow velocity increases. Since it was observed in Fig.~\ref{fig:8} that the stress is underestimated in the SAB model, we apply a numerical correction to the yield stress ($\Sigma_c^\text{SAB} = 0.87~\!\Sigma_c$) for comparison purpose, so that the first breaking event occurs at the same flow velocity.
The evolution of the drag is shown in Fig.~\ref{fig:10}, for $n=200$ in the SAB model.
We observe a very good agreement between SAB and FEM on the drag evolution under increasing flow velocity. The SAB model reproduces accurately the process of flow-induced pruning in a tree: the first drag reduction is well captured, as well as the subsequent slow increase.
In this figure, sudden drops of the drag correspond to breaking events.
The geometries  corresponding to $U=0.4$ and $U=1.5$ (in arbitrary units) show similar broken regions and thus good qualitative agreement. 
This second application of the space-averaged branching model validates this model in a more complex and realistic situation.

\section{Discussion and conclusions}
\label{sec:ConclusionAndDiscussion}

Interested in modeling conservation laws in ramified system in a continuous way, we developed in this work an original model for branched systems using space averaging, named space-averaged branching (SAB). 
Using standard averaging methods on branched systems described by one-dimensional equations, we  were able to derive a continuous formulation of conservation laws. We then used a first order approximation, reducing the model equations to a set of one-dimensional equations along characteristic curves $\car$, defined as streamlines of the space-averaged branch direction vector field $\vec{t}_\ss$. 
The SAB equivalent medium therefore is described by a set of characteristic curves, with one-dimensional balance equations identical to the initial ones, {except} for an additional term that contains the local information on the geometry. The derivation of this formulation only requires that the initial problem has a one-dimensional description along the segments of the system.
The resulting medium is not ramified, allowing for much simpler and faster computations.
The fields needed in the SAB model are essentially the average branch direction, and the volume fraction occupied by the branched system.

The additional term, $\nabla.(\solfrac \vec{t}_\ss)$, referred to as geometrical forcing, is the key element of the SAB model, as it accounts for the different geometrical features of the initial structure, including branching, which was the fundamental issue {in} the initial problem. 
This term, defined everywhere continuously, accounts for the discrete aspect of the real system, where each segment has to be treated separately with boundary conditions given by {branching point} relations. The SAB model therefore provides a continuous and uniform description in the volume occupied by the system.

We validated the space-averaged branching model by considering two different problems. The first one was a 1D pipe system, easily solved analytically. The SAB model was used to compute the flow rate along the system, and showed excellent agreement with the analytical solution. The problem, presented in the absence of external forcing, put a particular emphasis on the crucial role played by the geometrical forcing. 
The second problem considered for validation was the computation of mechanical loads induced by a static uniform wind on an idealized tree. The results were compared to finite elements computations, showing again a very good accuracy of the SAB model on the load distribution. We then considered the sequence of flow-induced pruning, that is the successive breakage of branches under an increasing wind velocity. The results, consistent with finite elements predictions, showed a fast convergence of the SAB computations in terms of number of characteristic curves. In particular, the convergence was found for a number of curves lower than 10\% of the maximum number of parallel branches in the tree.

The model was derived considering a first order approximation on {fluctuations compared to volume-averaged values. In particular, we showed that the ramification angle should be small. 
However, the example of wind-induced loads on trees showed that the necessary assumption is not very constraining.}
In fact, when the branching angle is reduced to 0 (not shown here for the sake of brevity), the maximum bending stress in the structure matches the finite elements prediction. The stress underestimation in the case $\brangle = 25^\circ$ can be explained by the large branching angle.
This may not be the only reason, the choice made for computing the bending stress is in fact critical in this case due to the diameter dependence. Despite this underestimation {of} the stress, the SAB model predicted accurately the load profiles and consistent values without any fitting parameter. Moreover, considering a random tree, it was shown that there is no particular symmetry required in the real system.

The model was validated considering particular geometries where any branching point is characterized by one incoming and several outgoing segments. In fact, the model was derived in the general case, and could be used for more complicated networks. For example, the SAB model could be applied to blood circulation, where entangled networks of capillaries are more easily described as a continuous medium.
{Moreover, the model was derived in the general case, and complex three-dimensional structures and systems could be modeled in a similar manner.}
The pipe example has natural applications, for sediment or pollutant transport, in rivers or artificial pipe networks.
For the problem of wind-induced loads on a tree, we presented here the case of a uniform cross flow, but this model can be used in a broader scope. 
The coupling of this model with a standard flow model should give additional insights on the interaction between a flow and vegetation. Models used for predicting wind damage to forests could be coupled with this homogeneous description, allowing {the possibility of describing} the type of damage that would occur in a forest {during} a storm. {So far, continuous forest models do not account for branching in trees, and consider either homogeneous media or arrays of cylinders as forest models \citep{gardineretal_2000, liu_2010}. }
As the presented case was derived under a uniform flow, we did not need to change the SAB fields, and could treat each characteristic curve independently. However, if flow modifications were to be significant, an actualization of these fields {would} become necessary.

{The space-averaged branching model developed in this paper can be used for describing a large variety of branched systems. The same methodology can be applied for obtaining continuous formulations of conservation laws in complex oriented networks, like internal flows in a pipe network, or in fact the transport of any quantity through a branched network. }

\appendix
\section{Finite volume related issues}
\label{sec:app_hg}

Introducing the finite volume of the system yields some technical issues at branching regions and at the borders of the averaging volume. We present here how these issues are overcome.

\begin{figure}[tb]
	\centering
		\includegraphics{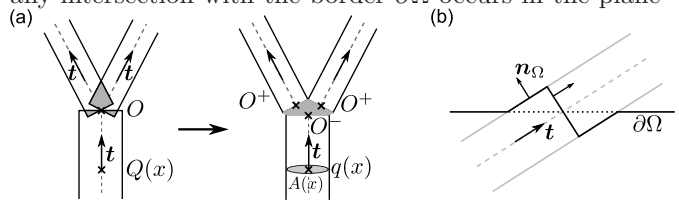}
	\caption{Averaging notations and hypothesis: (a) branching point representation; (b) adapted averaging volume border $\partial \vol$ in solid line, compared to the initial border in dotted line.}
	\label{fig:11}
\end{figure}

A first singularity appears at the branching points, concerning the definition of the continuous quantity $q(x)$, Fig.~\ref{fig:11}a. In order to overcome that singularity, the branching point is extended to a region whose typical length scale is the diameter of the segments $D$. Over this region the branching relations are conserved between $O^-$ and $O^+$ (see Fig.~\ref{fig:11}a). The function $q$ is therefore unknown in the branching region. However, this region has a volume that scales as $D^3$, and $D\ll L$ (the segment length), due to the branch slenderness. The length scale of $\vol$ being at least of the order of $L$, the volume represented by the branching regions is negligible compared to the solid volume $\vs$ and the total volume $\vol$.

A second issue is related to the intersection with the boundaries of the averaging volume. Considering that the branches have a non-zero volume is consistent only if any intersection with the border $\partial\Omega$ occurs in the plane normal to the tangent vector $\vec{t}$, otherwise, $q$ can take different values at the intersection. 
This is sketched in Fig.~\ref{fig:11}b, which shows this artefact of the three-dimensionalisation occurring at the borders $\partial \vol$ of the averaging volume. 
This issue is bypassed considering a modified border, shown in solid line in the sketch of Fig.~\ref{fig:11}b. The modification in volume that this adaptability will add is again negligible compared to the total volume $\vol$, since it scales as $D^2L$.

%
%

\end{document}